\documentclass[sigconf]{acmart}
\usepackage[ruled,vlined]{algorithm2e}
\usepackage{gensymb}
\usepackage{subcaption}
\usepackage{multirow}
\usepackage{array}

\definecolor{Red}{rgb}{1,0,0}
\definecolor{Blue}{rgb}{0,0,1}
\definecolor{Green}{rgb}{0,0.5,0}
\definecolor{Gray}{gray}{0.4}

\newcommand{\blue}[1]{\textcolor{Blue}{#1}}

\newcommand{\oursystem}{COHORT}
\AtBeginDocument{%
  \providecommand\BibTeX{{%
    \normalfont B\kern-0.5em{\scshape i\kern-0.25em b}\kern-0.8em\TeX}}}

\copyrightyear{2020}
\acmYear{2020}
\setcopyright{acmcopyright}
\acmConference[BuildSys '20]{The 7th ACM International Conference on Systems for Energy-Efficient Buildings, Cities, and Transportation}{November 16-19, 2020}{Yokohama, Japan}
\acmBooktitle{The 7th ACM International Conference on Systems for Energy-Efficient Buildings, Cities, and Transportation (BuildSys '20), November 16-19, 2020, Yokohama, Japan}
\acmPrice{15.00}
\acmDOI{10.1145/3408308.3427980}
\acmISBN{978-1-4503-8061-4/20/11}
\setcopyright{rightsretained} 

\begin{document}

\title{\oursystem: Coordination of Heterogeneous  Thermostatically Controlled Loads for Demand Flexibility}

\author{Bingqing Chen}
\orcid{1234-5678-9012}
\affiliation{%
  \institution{Carnegie Mellon University}
  \streetaddress{5000 Forbes Avenue}
  \city{Pittsburgh}
  \state{PA}
  \country{USA}}
\email{bingqinc@andrew.cmu.edu}

\author{Jonathan Francis}
\orcid{0000-0002-0556-1136}
\affiliation{%
  \institution{Carnegie Mellon University}
  \streetaddress{5000 Forbes Avenue}
  \city{Pittsburgh}
  \state{PA}
  \country{USA}}
\email{jmf1@cs.cmu.edu}

\author{Marco Pritoni}
\affiliation{%
  \institution{Lawrence Berkeley National Lab}
  \streetaddress{1 Cyclotron Rd}
  \city{Berkeley}
  \state{CA}
  \country{USA}}
\email{mpritoni@lbl.gov}
\orcid{1234-5678-9012}

\author{Soummya Kar}
\affiliation{%
  \institution{Carnegie Mellon University}
  \streetaddress{5000 Forbes Avenue}
  \city{Pittsburgh}
  \state{PA}
  \country{USA}}
\email{soummyak@andrew.cmu.edu}
\orcid{1234-5678-9012}

\author{Mario Berg\'{e}s}
\affiliation{%
  \institution{Carnegie Mellon University}
  \streetaddress{5000 Forbes Avenue}
  \city{Pittsburgh}
  \state{PA}
  \country{USA}}
\email{mberges@andrew.cmu.edu}
\orcid{1234-5678-9012}

\renewcommand{\shortauthors}{Chen, et al.}

\begin{abstract}
Demand flexibility is increasingly important for power grids. Careful coordination of thermostatically controlled loads (TCLs) can modulate energy demand, decrease operating costs, and increase grid resiliency. 
We propose a novel distributed control framework for the Coordination Of HeterOgeneous Residential Thermostatically controlled loads (COHORT). COHORT is a practical, scalable, and versatile solution that coordinates a population of TCLs to jointly optimize a grid-level objective, while satisfying each TCL's end-use requirements and operational constraints. To achieve that, we decompose the grid-scale problem into subproblems and coordinate their solutions to find the global optimum using the alternating direction method of multipliers (ADMM). The TCLs' local problems are distributed to and computed in parallel at each TCL, making COHORT highly scalable and privacy-preserving. While each TCL poses combinatorial and non-convex constraints, we characterize these constraints as a convex set through relaxation, thereby making \oursystem~computationally viable over long planning horizons. After coordination, each TCL is responsible for its own control and tracks the agreed-upon power trajectory with its preferred strategy. In this work, we translate continuous power back to discrete \textit{on}/\textit{off} actuation, using pulse width modulation. COHORT is generalizable to a wide range of grid objectives, which we demonstrate through three distinct use cases: generation following, minimizing ramping, and peak load curtailment. In a notable experiment, we validated our approach through a hardware-in-the-loop simulation, including a real-world air conditioner (AC) controlled via a smart thermostat, and simulated instances of ACs modeled after real-world data traces. During the 15-day experimental period, COHORT reduced daily peak loads by an average of 12.5\% and maintained comfortable temperatures.  
\end{abstract}



\maketitle

\section{Introduction}
Growing peak demand in some regions and increasing penetration of renewable generation in others are presenting challenges for grid operators to balance supply and demand \cite{eiapeak, duckcurve2017}.
Traditionally, demand-side load is viewed as uncontrollable, while supply-side resources manage power generation to match it. An emerging paradigm is to tap into the flexibility of demand-side resources to reduce, shift, or modulate their loads in response to price or control signals \cite{neukomm2019grid}. Such demand flexibility can be utilized to provide grid services, improve grid resiliency \cite{neukomm2019grid}, and reduce operating costs \cite{callaway2010achieving}. 

In this work, we focus on load control\footnote{As opposed to methods based on economic incentives. We refer interested readers to \cite{callaway2010achieving} for a comparison of load control vs. price-based methods. }\cite{callaway2010achieving} of residential thermostatically controlled loads (TCLs), such as air conditioners (ACs), refrigerators, and electric water heaters, which account for about 20\% of the electricity consumption in the United States (US) \cite{hao2014aggregate}. Due to their inherent flexibility through thermal inertia, they can provide grid services without compromising their end uses.
However, there are two challenges to utilizing TCL flexibility. Firstly, TCLs must be aggregated across a population to be a meaningful resource at the grid-level \cite{neukomm2019grid}, which results in a control problem with a large state-action space. Secondly, the constraints posed by each TCL are combinatorial and thus non-convex \cite{burger2017generation}, due to the fact that a TCL operates in discrete action space, i.e., \textit{on} or \textit{off}.

\begin{figure}[]
  \includegraphics[width=\linewidth]{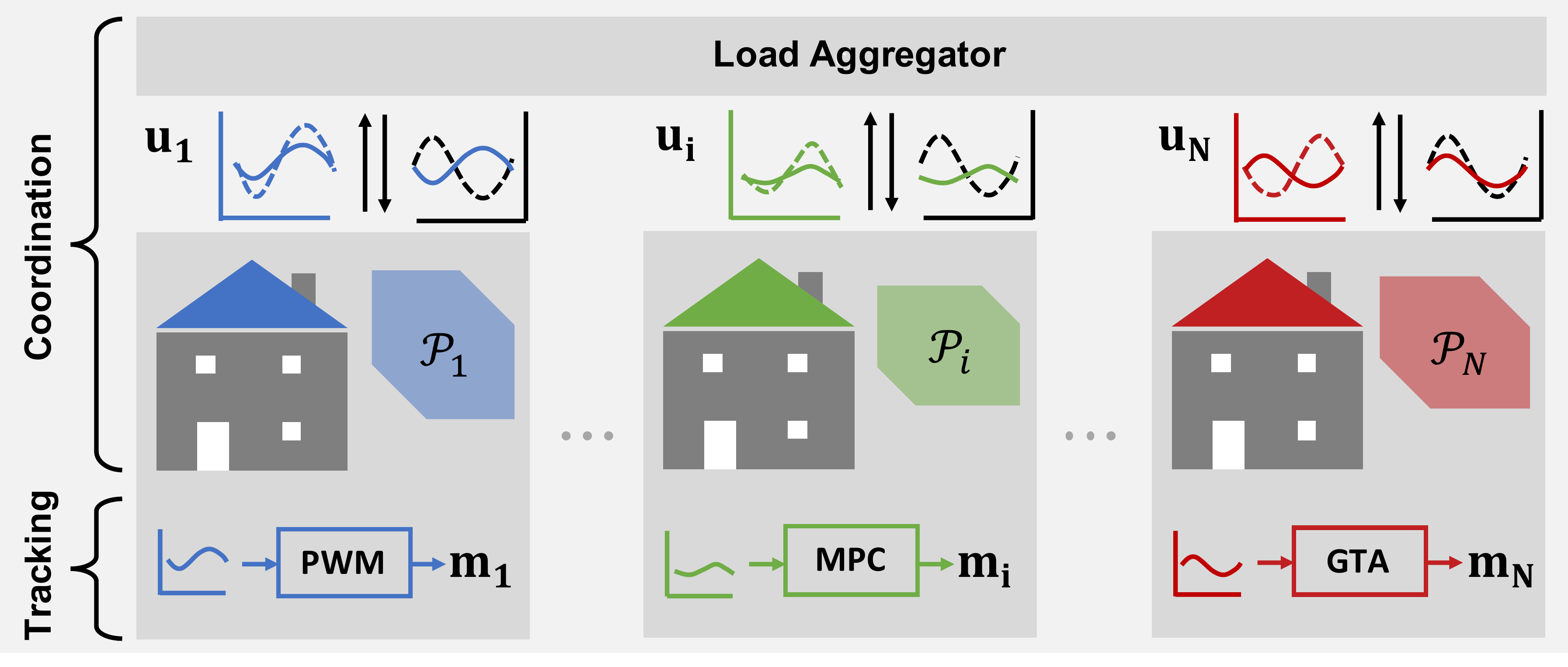}
  \caption{Framework. The load aggregator coordinates a population of TCLs to jointly optimize a grid-level objective, while satisfying each TCL's constraints, characterized by the set $\mathcal{P}_i$.
  The load aggregator and each TCL coordinates at the level of its power trajectory, $\mathbf{u_i}$, until a consensus is reached among the population. Each TCL is responsible for its own control and tracks $\mathbf{u_i}$ locally with its preferred strategy. }
   \label{fig:framework}
\end{figure}

We present a novel framework (Figure \ref{fig:framework}) for the Coordination Of HeterOgeneous Residential Thermostatically controlled loads (\oursystem) to jointly optimize a grid-level objective, while satisfying each TCL’s end-use requirements and operational constraints. To effectively handle the large state-action space, we adopt a distributed control architecture, where each TCL is responsible for its own control and coordinates with others to find a grid-level solution.
Similar to \cite{burger2017generation}, we decompose the grid-level problem into smaller subproblems and coordinate their solutions using the alternating direction method of multipliers (ADMM) \cite{boyd2011distributed}. The advantages of the distributed architecture compared with centralized and decentralized approaches are elaborated in Section \ref{sec:review}. 
To address the second challenge, we characterize each TCL's flexibility, i.e., the set of all admissible power profiles \cite{zhao2017geometric}, as a convex set through relaxation. As a result, \oursystem~is computationally viable for tasks with long planning horizons (e.g., 24 hours), thereby addressing a limitation of  \cite{burger2017generation}---namely that its computational cost grows exponentially with the planning horizon. After coordination, we use low-frequency pulse width modulation (PWM), inspired by \cite{burke2009low}, to translate the solution of the convex-relaxed problem back to \textit{on}/\textit{off} activation. Since the coordination process makes no assumption on each TCL's control scheme, the TCL may opt for alternative strategies with reference-tracking capability, such as model predictive control (MPC) and global thermostat adjustment (GTA).

 


\oursystem~can incorporate detailed, system-specific dynamics and constraints of individual TCLs. At the same time, its computational cost scales well with both population size and planning horizon. As a result, \oursystem~is generalizable to a wider variety of grid objectives, compared to existing methods, which we demonstrate through three distinct use cases: generation following, minimizing ramping, and peak load curtailment. As a proof of concept, we evaluated \oursystem~in simulation, targeting challenges arising from increasing penetration of renewable generation \cite{duckcurve2017} under, generation following and minimizing ramping, based on load profiles from California Independent System Operator (CAISO) \cite{CAISO2020}.  

Then, we validated that  \oursystem~is practical for real-world TCLs. To do that, we developed a hardware-in-the-loop (HIL) simulation, including a real-world residential AC controlled via a smart thermostat, and simulated instances of ACs modeled after real-world data traces. In a 15-day experimental period, \oursystem~ reduced daily peak loads by 12.5\% on average based on load profiles from Pennsylvania-New Jersey-Maryland (PJM) Interconnection \cite{PJM2018}, while maintaining comfort in the real-world testbed. 





The rest of the paper is organized as follows. Section \ref{sec:review} reviews prior work on TCL control. Section \ref{sec:background} provides technical background, and Section~\ref{sec:approach} formally introduces \oursystem. We then describe experiments and results both in the simulation study (Section~\ref{sec:exp-1}) and in the HIL simulation (Section~\ref{sec:exp-2}), and present conclusions and possible future work in Section~\ref{sec:discussion}.

\section{Related Work} \label{sec:review}
We review existing work on TCL control, and include work on other flexible loads if the methodology is relevant. 
\paragraph{Architectures for TCL control}
The primary challenge for jointly controlling a large number of TCLs is the large state-action space. To address this challenge, a popular approach is to develop an aggregate model for the population and control the population in a centralized manner. Examples of such aggregate model include the state bin transition model \cite{koch2011modeling, zhang2013aggregated} and the virtual battery model \cite{hao2014aggregate, zhao2017geometric}. However, these aggregate models depend on the assumptions that each system may be characterized by a 1\textsuperscript{st}- (or 2\textsuperscript{nd}- \cite{zhang2013aggregated}) order linear model, and that all systems in the population share the same model structure and control scheme. These aggregate models have low fidelity and do not capture system-specific dynamics. Specifically, aggregate TCL modeling is ill-suited for predicting long-term responses---a pre-requisite for tasks with long planning horizons, such as load shifting \cite{burger2017generation}. Alternatively, one can jointly control building loads as a centralized MPC problem \cite{vrettos2016robust}, but, while this approach allows for incorporation of detailed building models and system-specific constraints, it is computationally expensive and also raises privacy concers, as it requires each building to share an excessive amount of information with the load aggregator, including: thermal models, system specifications, control logic, and occupants' usage pattern and comfort preferences. 

Aside from the centralized architecture, decentralized control \cite{tindemans2015decentralized} and distributed control \cite{costanzo2013coordination, mahdavi2017model, burger2017generation} approaches have also been proposed in the literature. Taking advantage of the fact that system frequency is a universally-available indicator for supply-demand imbalance, \cite{tindemans2015decentralized} determines the action of each TCL with a power response model based on locally-available information. The key advantage of a decentralized control approach is that each system can be controlled based on local information, without any communication. However, this characteristic also constrains the applications of decentralized control to frequency or voltage regulation and real-time load shaping \cite{burger2017generation}. 

In a distributed architecture, which we adopt in this work, each system is responsible for its own control, and it coordinates with others to jointly achieve a grid-level objective. 
In \cite{mahdavi2017model}, the distributed MPC scheme allocated the aggregate load to TCL clusters following a time-invariant weight. This allocation scheme does not account for the fact that the flexibility available at each building is time-varying, and thus does not fully utilize the aggregate flexibility \cite{vrettos2016robust}.  \cite{Chen2020Learning} used a distributed MPC approach similar to \cite{mahdavi2017model}, but learned the allocation scheme with an evolutionary strategy. Most similar to this work is \cite{burger2017generation}, which also used ADMM for distributed optimization. A major advantage is that the computation is distributed to and parallelized at each TCL. A significant limitation of \cite{burger2017generation} the computational cost grows exponentially with the planning horizon. Specifically, it represented each TCL as a set of feasible state-action trajectories, the size of which is $N_a^T$, where $N_a$ is the number of alternative actions considered at each time-step, and T is the number of time-steps in the planning horizon. Given that the trajectories depends on initial state and future disturbances, the trajectories need to be unrolled at each time-step. Another limitation is that the solution is only optimal with respect to the set of trajectories under consideration. 

\paragraph{Experimental Validation} The majority of works on this topic validated their approaches in simulation. In particular, works such as \cite{koch2011modeling, hao2014aggregate, tindemans2015decentralized, zhao2017geometric, mahdavi2017model, burger2017generation} validated their approach on population of TCLs simulated with 1\textsuperscript{st}-order linear thermal model, using model parameters sampled from assumed distributions. It is unclear how well such validation reflects performance on real-world systems. \cite{xu2014modeling} demonstrated 1\textsuperscript{st}- and 2\textsuperscript{nd}-order models failed to accurately capture the thermodynamics of an individual electric water heater. Furthermore, there is a large difference between performance reported in simulation and in real-world testbeds. For instance, \cite{hao2014aggregate} reported a maximum absolute percentage error of less than 1\% tracking a 4s frequency regulation signal in a simulation study. In comparison, in a real-world experiment on 300 residential ACs, \cite{muller2019large} reported a median absolution percentage error of 6.7\% executing 1-hour demand response (DR) events, which is arguably a much simpler task. Such discrepancy calls for more realistic evaluation. Other attempts at realistic evaluation include \cite{cole2014community} which developed linear models based on configurations of real households, and \cite{nghiem2017data}, which used a co-simulation environment with EnergyPlus models. 

%


\paragraph{Optimization Objectives}
A myriad of grid-level objectives have been discussed in the literature, such as:
energy cost minimization \cite{cole2014community}, DR events \cite{zhang2013aggregated, tindemans2015decentralized, nghiem2017data, muller2019large}, frequency regulation \cite{hao2014aggregate, zhao2017geometric}, generation following \cite{mahdavi2017model, burger2017generation}, reference tracking \cite{koch2011modeling}, and peak load reduction \cite{costanzo2013coordination}. 
However, these works generally formulate their approaches based on their specific use case, without discussing their generalizability to other applications. 


\section{Background}\label{sec:background}
We now present background technical concepts used by \oursystem, including TCL modeling (Section \ref{sec:TCL}) and ADMM (Section \ref{sec:admm}).

\subsection{TCL Model and Flexibility}\label{sec:TCL}
Here, we introduce the modeling of an individual TCL and define its flexibility. The contents is largely inspired by \cite{zhao2017geometric}, from which we made modifications based on our problem.

\subsubsection{System Dynamics}
The temperature dynamics of an individual TCL is commonly modeled with Eq. \ref{eq:sys_dynamics} \cite{koch2011modeling, hao2014aggregate, zhao2017geometric}, where $T_t$ is the TCL temperature, $T_{a,t}$ is the ambient temperature, and $m_t\in\{0, 1\}$ is the binary control variable representing the operating state, i.e., \textit{on} or \textit{off}, at time $t$. The negative sign associated with the control action assumes the TCL is operating in cooling mode, which could be changed to a positive sign to reflect heating. $P_m$ is the rated power. Denoting the thermal resistance and capacitance of the TCL as $R$ and $C$ respectively, the model parameters can be calculated as: $a=\exp\{-\Delta T/(RC)\}$ and $b_t = \eta_t R$, where $\Delta T$ is the time-step and $\eta_t$ is the time-varying coefficient of performance (COP). While the thermodynamics is linear, it is difficult to analyze the system dynamics in Eq. \ref{eq:sys_dynamics} due to the discrete control variable $m_t$. A common approach is to apply convex relaxation to the discrete control variable, which results in a linear system approximation (Eq. \ref{eq:sys_relaxed}) \cite{koch2011modeling, hao2014aggregate, zhao2017geometric}. The new control variable $u_t\in[0, P_m]$, i.e., power consumption of the TCL, is continuous instead of discrete. This approach is justified as the aggregate behavior of a TCL population can be approximated accurately by Eq. \ref{eq:sys_relaxed} \cite{zhao2017geometric}.
\begin{subequations}
\begin{equation}\label{eq:sys_dynamics}
    T_{t+1} = a T_t + (1-a)(T_{a,t}-b_t m_t P_m)
\end{equation}
\begin{equation}\label{eq:sys_relaxed}
    T_{t+1} = a T_t + (1-a)(T_{a,t}-b_t u_t)
\end{equation}
\end{subequations}

The TCL dynamics over a planning horizon, $t:t+T-1$, is thus characterized by Eq. \ref{eq:vb_evolve} (or more concisely Eq. \ref{eq:vb_matrix}), where $\mathbf{B_u}= \text{diag}(-(1-a)b_t, \dots, -(1-a)b_{t+T-1})$. Throughout this work, boldface lower-case letters, e.g., $\mathbf{x}$, are vectors, and boldface upper-case letters, e.g., $\mathbf{A}$, are matrices; $\mathbf{x},\; \mathbf{x}_0, \;\mathbf{u}\in {\rm I\!R}^T$ and $\mathbf{A},\; \mathbf{B_u}\in {\rm I\!R}^{T\times T}$. Denoting the number of disturbance terms as $l$, we have $D\in {\rm I\!R}^{T\times l}$ and $\mathbf{b_d}\in {\rm I\!R}^{l}$. In this case, the disturbance term only includes the ambient temperature, $l=1$.
\begin{subequations}
\begin{equation}
\label{eq:vb_evolve}
\underbrace{\begin{bmatrix}
    1  &  &  &  \\
    -a & 1 & & \\
       & \ddots   &\ddots & \\
     &    &   -a  &  1
\end{bmatrix}}_{\mathbf{A}}
\underbrace{\begin{bmatrix}
T_{t+1}\\
T_{t+2}\\
\vdots\\
T_{t+T}
\end{bmatrix}}_{\mathbf{x}}
=\underbrace{\begin{bmatrix}
aT_{t}\\
0\\
\vdots\\
0
\end{bmatrix}}_{\mathbf{x}_0}
+
\mathbf{B_u}
\underbrace{\begin{bmatrix}
u_{t}\\
u_{t+1}\\
\vdots\\
u_{t+T-1}
\end{bmatrix}}_{\mathbf{u}}
+
\underbrace{\begin{bmatrix}
T_{a,t}\\
T_{a,t+1}\\
\vdots\\
T_{a,t+T-1}
\end{bmatrix}}_{\mathbf{D}}\underbrace{\begin{bmatrix}
1-a
\end{bmatrix}}_{\mathbf{b_d}}
\end{equation}
\begin{equation}\label{eq:vb_matrix}
    \mathbf{A}\mathbf{x} = \mathbf{x}_0+\mathbf{B_u}\mathbf{u}+\mathbf{D}\mathbf{b_d}
\end{equation}
\end{subequations}
\paragraph{Constraints} Each TCL needs to satisfy the end-use requirements and respect the operational constraints. In this case, we require the TCL temperature to be within the deadband, i.e., $T_t\in[T_{sp}-\Delta,T_{sp}+\Delta]$, where $T_{sp}$ is the setpoint and $\Delta$ is half of the deadband. At the same time, the system needs to be operating with in its power limits, i.e., $P_t \in [0, P_m]$. 
Combining the system dynamics given in Eq. \ref{eq:vb_matrix}, the aforementioned constraints can be written as Eq. \ref{eq:constraints}, where $\underline{\mathbf{u}} = [0]$, $\overline{\mathbf{u}} = [Pm]$, $\underline{\mathbf{x}} = [T_{sp}-\Delta]$, and $\overline{\mathbf{x}} = [T_{sp}+\Delta]$.
\begin{equation} \label{eq:constraints}
\underline{\mathbf{u}} \leq \mathbf{u} \leq \overline{\mathbf{u}}; \quad \underline{\mathbf{x}} \leq \mathbf{A^{-1}}( \mathbf{x}_0 + \mathbf{B_u} \mathbf{u} + \mathbf{D} \mathbf{b_d} ) \leq \overline{\mathbf{x}};
\end{equation}

\subsubsection{Flexibility} We adopt the definition of flexibility proposed in \cite{zhao2017geometric}, where the flexibility of a system is the set of all admissible power profiles. Our formulation of Eq. \ref{eq:def_flex} is generalized from that in \cite{zhao2017geometric}, to incorporate non-linear systems. 
$\mathcal{P}$ denotes the flexibility of the system, $\mathcal{T}(x_{k}, u_k)$ denotes the state transition function, and $\underline{x}_k$, $\overline{x}_k$, $\underline{u}_k$, and $\overline{u}_k$ are the lower and upper bounds for state and action of the given system at time $k$. An important intuition is that the flexibility is coupled over time through the thermodynamics \cite{zhao2017geometric}: 
\begin{equation}\label{eq:def_flex} 
    \mathcal{P}= \left\{[u_{t:t+T-1}]
    \left\vert \begin{array}{l}
    x_{k+1} = \mathcal{T}(x_{k}, u_k);\\
    \underline{u}_k\leq u_k\leq \overline{u}_k; \;\;\forall k \in \{t, \dots , t+T-1\} \\
    \underline{x}_{k+1}\leq x_{k+1}\leq \overline{x}_{k+1};
  
    \end{array}\right.
    \right\}
\end{equation}

For the specific case of TCL, which follows the linear dynamics in Eq. \ref{eq:sys_relaxed}, the flexibility $\mathcal{P}$ of a TCL can be expressed as Eq. \ref{eq:flex}. 
\begin{equation}\label{eq:flex}
    \mathcal{P} = \mathcal{U} \cap \mathcal{X} 
\end{equation}
where $\mathcal{U}=\{\mathbf{u} | \underline{\mathbf{u}} \leq \mathbf{u} \leq \overline{\mathbf{u}}\}$ and $\mathcal{X}=\{\mathbf{x} |\underline{\mathbf{x}} \leq \mathbf{A^{-1}}( \mathbf{x}_0 + \mathbf{B_u} \mathbf{u} + \mathbf{D} \mathbf{b_d} ) \leq \overline{\mathbf{x}}\}$, as derived in Eq. \ref{eq:constraints}. Note that $\mathcal{P}$ boils down to a set of linear inequalities, which is geometrically interpreted as a polytope\footnote{A polytope can be characterized as a set $\mathcal{S}=\{x\in \mathbb{R}^n|Ax\leq b\}$.} \cite{zhao2017geometric}.

\subsection{ADMM}\label{sec:admm}

ADMM is a well-established distributed convex optimization algorithm, which decomposes a large problem into smaller subproblems and coordinates the solutions to find a global optimum \cite{boyd2011distributed}. Generally, ADMM solves problems in the form of Eq. \ref{eq:ADMM}. 
\begin{equation}\label{eq:ADMM}
\begin{aligned}
\min_{u, v} \quad &  f(u)+g(v)\\
\textrm{s.t.} \quad & Au+Bv=c
\end{aligned}
\end{equation}
Specifically, we introduce the application of ADMM to a canonical problem: the \textit{sharing problem}, as given in Eq. \ref{eq:sharing}, where $f_i$ is a local objective for agent $i$, and $g$ is the global objective---defined as a function of the aggregate of all decision variables from the agents:

\begin{equation}\label{eq:sharing}
\begin{aligned}
\min_{u_i} \quad &  \sum_i^N f_i(u_i)+g(\sum_i^N u_i)
\end{aligned}
\end{equation}

By introducing a copy of the decision variable $u_i$ as $v_i$, the sharing problem can be written in a ADMM-compatible form (Eq. \ref{eq:ADMM_sharing}):

\begin{equation}\label{eq:ADMM_sharing}
\begin{aligned}
\min_{u, v} \quad &  \sum_i^N f_i(u_i)+g(\sum_i^N v_i)\\
\textrm{s.t.} \quad & u_i -v_i = 0,\quad i = 1, \dots, N
\end{aligned}
\end{equation}

The update rules for solving the problem are given in Eq. \ref{eq:updates}, where $w$ is the dual variable, $\rho$ is a hyperparameter, and the superscript $(k)$ denotes the value of a variable at the k\textsuperscript{th} iteration. We elaborate on the intuition behind these update rules here. The sharing problem can be interpreted as the agents coordinating their decisions so as to strike a balance between the local and the global objectives. Hence, $u_i$ and $v_i$ are each agent's solutions to its local problem and the global objective, respectively. The dual variable $w_i$, as calculated in Eq. \ref{eq:w-update}, is the cumulative disagreement between $u_i$ and $v_i$. Thus, $w_i$, which we also call the incentive variable, communicates how to adjust each agent's solutions such that they would agree, i.e., $u_i=v_i$. Thus, in the \textit{u-update} step (Eq. \ref{eq:u-update}), each agent solves its local problem, while mindful of its solution to the global problem. Similarly, in the \textit{v-update} step (Eq. \ref{eq:v-update}), the agents jointly optimize the global objective, while ensuring their decisions are close to those of their local problems. 


\begin{subequations}\label{eq:updates}
\begin{align}
u_i^{(k+1)} =& \arg \min_{u_i} f_i(u_i)+\frac{\rho}{2}\| u_i -v_i^{(k)}+w_i^{(k)}\|^2_2\label{eq:u-update}\\
v_i^{(k+1)} =& \arg \min_{v_i} g(\sum_i^N v_i) + \frac{N\rho}{2}\|u_i^{(k+1)} -v_i+{w_i}^{(k)}\|^2_2\label{eq:v-update}\\
{w}_i^{(k+1)} =& {w_i}^{(k)} + u_i^{(k+1)}-v_i^{(k+1)}\label{eq:w-update}
\end{align}
\end{subequations}

A downside of the updates rules given in Eq. \ref{eq:updates} is that it requires a copy of the decision variable for each agent, i.e., $v_i$. Intuitively, the global objective only depends on the aggregate behavior of the population, and thus a more efficient algorithm (Eq. \ref{eq:averaged-updates}) is possible using the mean of the variables, denoted as $\bar{u}$, $\bar{v}$, and $\bar{w}$ respectively.

\begin{subequations}\label{eq:averaged-updates}
\begin{align}
u_i^{(k+1)} =& \arg \min_{u_i} f_i(u_i)+\frac{\rho}{2}\| u_i - u_i^{(k)}+\bar{u}^{(k)}-\bar{v}^{(k)}+\bar{w}^{(k)}\|^2_2\label{eq:averaged-u-update}\\
\bar{v}^{(k+1)} =& \arg \min_{\bar{v}} g(N\bar{v}) + \frac{N\rho}{2}\|\bar{u}^{(k+1)}-\bar{v}+\bar{w}^{(k)}\|^2_2\label{eq:averaged-v-update}\\
\bar{w}^{(k+1)} =& \bar{w}^{(k)} + \bar{u}^{(k+1)}-\bar{v}^{(k+1)}\label{eq:averaged-w-update}
\end{align}
\end{subequations}

An equality such as $v_i^{(k)} = \bar{v}^{(k)}+(u_i^{(k)}+w_i^{(k)})-(\bar{u}^{(k)}+\bar{w}^{(k)})$, to show Eq. \ref{eq:updates} and Eq. \ref{eq:averaged-updates} are equivalent, can be derived from the stationarity condition \cite{burger2017generation}. For more details on the algorithm, we refer interested readers to \cite{boyd2011distributed}.

\section{Approach} \label{sec:approach}
We first formulate the problem and elaborate on the optimization procedure, with focus on the coordination between the load aggregator and the TCLs (Section \ref{sec:prob_formulation}). We then describe a PWM-based strategy for TCL-level control (Section \ref{sec:pwm}).

\subsection{Problem Formulation and Optimization}\label{sec:prob_formulation}

The problem we address can be formulated as Eq. \ref{eq:prob}: we want to simultaneously optimize a grid-level objective $g(\cdot)$, which is a function of aggregate power consumption, and make sure the actions of each TCL are admissible based on operational constraints and end-use requirements, characterized by the set $\mathcal{P}_i$. By applying convex relaxation to discrete actions, and characterizing the flexibility as a convex set $\mathcal{P}_i$, our approach is computationally viable for tasks with long planning horizons. Recall the definition in Section \ref{sec:TCL}, $\mathbf{u_i}\in {\rm I\!R}^T$ is the power consumption of a TCL over a planning horizon. The subscript $i$ denotes the i\textsuperscript{th} TCL. While we did not include any TCL-level objective other than satisfying its constraints, it is possible to incorporate such objectives. 
\begin{equation}\label{eq:prob}
\begin{aligned}
\min_{\mathbf{u_i}} \quad & g(\sum_i^N \mathbf{u_i})\\
\textrm{s.t.} \quad & \mathbf{u_i}\in \mathcal{P}_i,\; \forall i
\end{aligned}
\end{equation}
The problem in Eq. \ref{eq:prob} can be written in a ADMM-compatible form (Eq. \ref{eq:prob_admm}) by (i) introducing a copy of the variable $\mathbf{u_i}$ as $\mathbf{v_i}$, and (ii) representing the constraints set $\mathcal{P}_i$ with the indicator function $\mathbb{I}_{\mathcal{P}_i}$. By definition, $\mathbb{I}_{\mathcal{P}_i}(\mathbf{u_i}) = 0$, if $\mathbf{u_i}\in \mathcal{P}_i$, else $\mathbb{I}_{\mathcal{P}_i}(\mathbf{u_i}) = \infty$ \cite{ryan_pgd}. The large penalty for an inadmissible $\mathbf{u_i}$ forces the solver to find $\mathbf{u_i}$ that satisfies the constraints. 

\begin{equation}\label{eq:prob_admm}
\begin{aligned}
\min_{\mathbf{u_i}, \mathbf{v_i}} \quad &  \sum_i \mathbb{I}_{\mathcal{P}^i}\big(\mathbf{u_i}\big)+g\big(\sum_i  \mathbf{v_i}\big)\\
\textrm{s.t.} \quad & \mathbf{u_i}= \mathbf{v_i}
\end{aligned}
\end{equation}

\begin{figure}
    \centering
    \includegraphics[width = \linewidth]{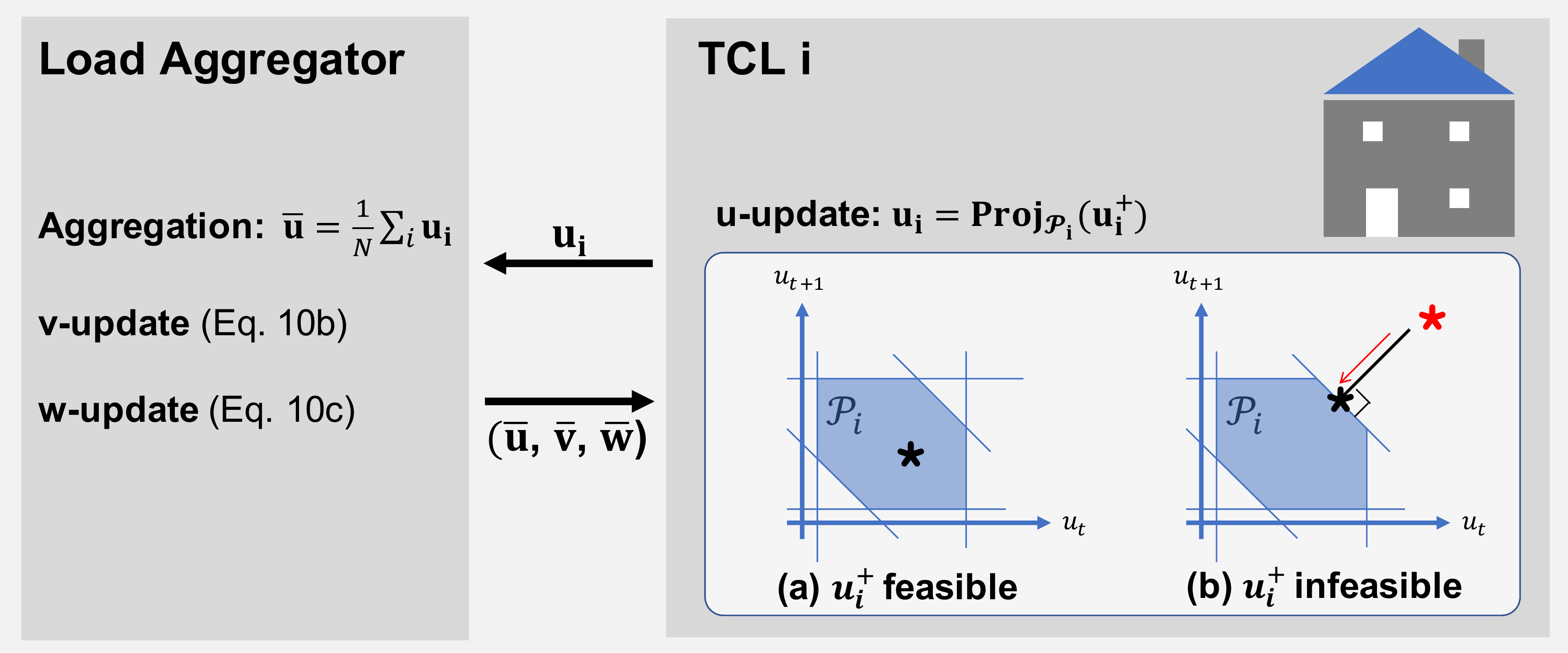}
    \caption{Coordination. The \textit{u-update} step is distributed to and computed in parallel at each TCL as a projection operation. The load aggregator calculates $\mathbf{\bar{u}}$, and then updates $\mathbf{\bar{v}}$ and $\mathbf{\bar{w}}$. Finally, the load aggregator broadcasts the $\mathbf{\bar{u}}$, $\mathbf{\bar{v}}$, and $\mathbf{\bar{w}}$ to the population. This procedure repeats until convergence.}
    \label{fig:coordination}
\end{figure}

Note that Eq. \ref{eq:prob_admm} now has the same form as Eq. \ref{eq:ADMM_sharing}, and thus may be solved with the update rules in Eq. \ref{eq:averaged-updates}.
Figure \ref{fig:coordination} summarizes the coordination procedure between the load aggregator and each TCL. Firstly, each TCL updates its action, $\mathbf{u_i}$, locally. 
Given that $f_i = \mathbb{I}_{\mathcal{P}_i}$ and $\mathcal{P}_i$ is a polytope for a TCL, the \textit{u-update} step (Eq. \ref{eq:averaged-u-update}) may be implemented efficiently as a projection operation \cite{ryan_pgd}, also illustrated in Figure \ref{fig:coordination}. To simplify notation, we denote $\mathbf{u_i}^+ =  \mathbf{v_i}^{(k)}-\mathbf{w_i}^{(k)}= \mathbf{u_i} ^{(k)}-\bar{\mathbf{u}}^{(k)}+\bar{\mathbf{v}}^{(k)}-{\mathbf{\bar{w}}}^{(k)}$; $\mathbf{u_i}^+$ can be interpreted as the desired power profile for TCL $i$ at the end of the k\textsuperscript{th} iteration, and the projection  $\mathbf{Proj}_{\mathcal{P}_i}(\mathbf{u_i}^+)$ ensures that the coordinated power profile is admissible for the TCL. 
Secondly, the load aggregator collects the actions from all agents to find the mean, $\mathbf{\bar{u}}$, and sequentially updates $\mathbf{\bar{v}}$ following Eq. \ref{eq:averaged-v-update}, and $\mathbf{\bar{w}}$ following Eq. \ref{eq:averaged-w-update}. Finally, the load aggregator broadcasts $\mathbf{\bar{u}}$, $\mathbf{\bar{v}}$, and $\mathbf{\bar{w}}$ to all TCLs, such that they can update $\mathbf{u_i}$ locally. This procedure repeats until convergence. 


While we use ADMM in this work, we note that alternative distributed optimization algorithms exist in the literature, such as consensus-based approaches \cite{hug2015consensus+} that are especially relevant for distributed energy management. However, for the application in this paper, ADMM is suitable given the communication topology\footnote{i.e. there being a load aggregator that centrally collects and broadcasts information. In comparison, consensus-based methods would be more suitable for peer-to-peer communication topology.} and provides scalability given the ease of solving each TCL's local problem.
There are three key advantages to the ADMM-based approach for this application. Firstly, the ADMM naturally decomposes the grid-level problem into subproblems. Thus, each TCL can ensure its local objective and constraints are satisfied, without sharing them with the load aggregator, thereby preserving privacy. Secondly, the \textit{u-update} at each TCL is computed in parallel. Thus, the approach is highly scalable to large population.  Finally, ADMM is guaranteed to converge to the grid-level optimum given a convex objective \cite{boyd2011distributed}. As will become clear through our demonstrations, a variety of grid objectives can be formulated as convex problems.

\subsection{TCL-level Control}\label{sec:pwm}

\begin{figure}
    \centering
    \includegraphics[width = \linewidth]{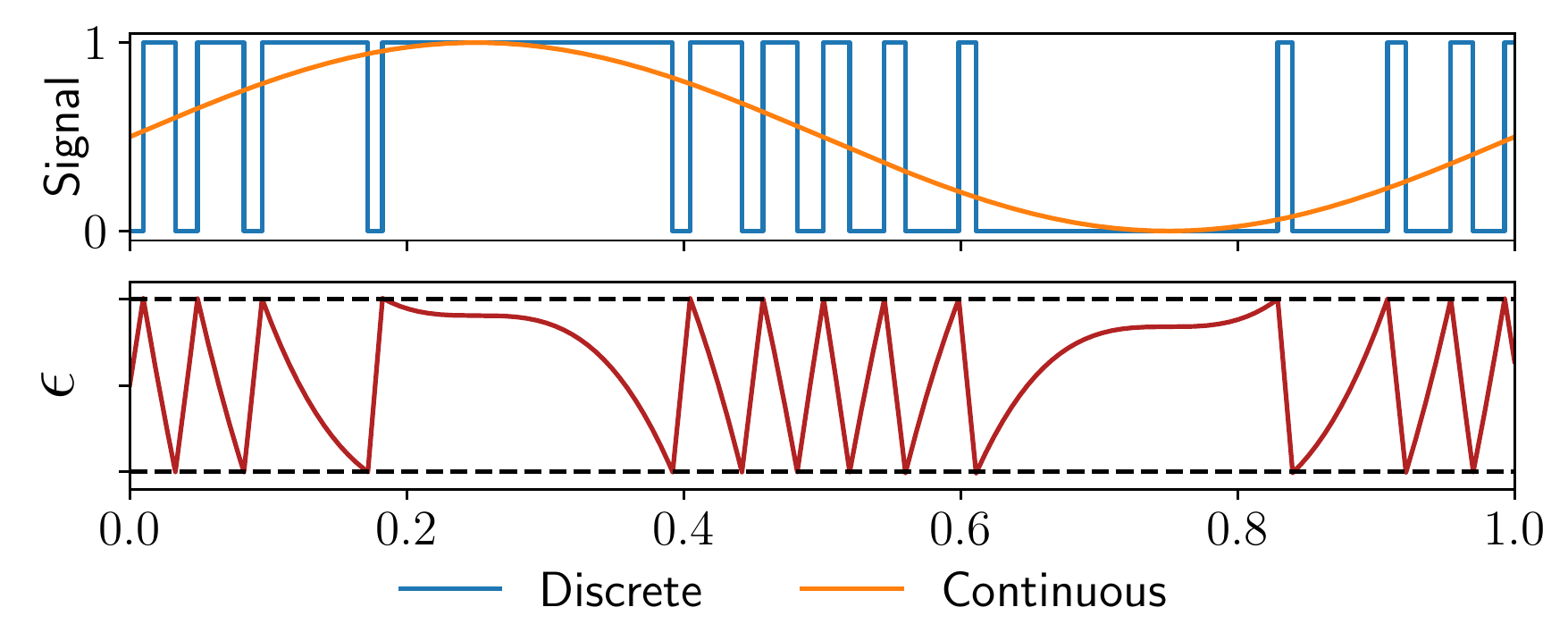}
    \caption{Sigma-Delta modulation converts a continuous signal to a discrete signal by switching states when the cumulative error, $\epsilon$, exceeds the limits (dashed black lines). }
    \label{fig:pwm_demo}
\end{figure}

Recall that we applied convex relaxation to TCL dynamics, i.e., $m_t P_m\in \{0, P_m\}$ (Eq. \ref{eq:sys_dynamics}) to $u_t\in [0, P_m]$ (Eq. \ref{eq:sys_relaxed}), such that the grid-level objective can be optimized efficiently over long time horizons. In this section, we describe how to translate the continuous power trajectory back to on/off actuation with PWM, a method for generating quasi-continuous output from an on/off actuator.
 In \cite{burke2009low}, it was demonstrated that a TCL could be treated as a variable power unit via low-frequency PWM. The action normalized by rated power, $u_t/P_m \in [0, 1]$, can be interpreted as duty cycle ratio, i.e., the portion of time the TCL is \textit{on} within the control time-step. Specifically, we implemented PWM with Sigma-Delta ($\Sigma$-$\Delta$) modulation \cite{park1999principles}, where TCL switches between \textit{on} and \textit{off} when the cumulative error, $\epsilon_t = \sum_{k=0}^t (u_k /P_m - m_k ) \Delta T$,  exceeds the limits. Figure \ref{fig:pwm_demo} illustrates its use in tracking a sine wave.

It is clear from Figure \ref{fig:pwm_demo} that fewer switchings are needed when a signal is close to either 0 or 1. We also observe that the solution tends to be close to a feasible initialization. The intuition may be that an individual TCL does not need to drastically change its default behavior to collaboratively achieve a grid-level objective. Given these observations, we initialize each $\mathbf{u_i}$ with a sequence of interlaced 0s and $P_m$s to encourage sparsity in the solution. Furthermore, by placing $0$s and $P_m$s with care in the initialization, short-cycling can be reduced. 

\section{Experiment 1: Simulation Study}\label{sec:exp-1}
In this section, we evaluated \oursystem~in simulation as an initial proof of concept. Following \cite{koch2011modeling, hao2014aggregate, tindemans2015decentralized, zhao2017geometric, mahdavi2017model, burger2017generation}, we validated our approach on a population of TCLs simulated with 1\textsuperscript{st}-order linear thermal models (Eq. \ref{eq:sys_dynamics}). We simulated 1000 TCLs, using parameters sampled from uniform distributions around nominal values \cite{koch2011modeling, hao2014aggregate,tindemans2015decentralized, zhao2017geometric}. Specifically, we followed the same parameter distributions and values for temperature setpoint and exogenous variable as \cite{hao2014aggregate}. We used a deadband of $\Delta$= 1\textsuperscript{o}C  throughout this work. While these assumptions may not reflect realistic system dynamics and population heterogeneity, we adopted them such that the performance of our approach is directly comparable to those reported in the literature. We lifted these assumption and validated \oursystem~in a real-world testbed in Section \ref{sec:exp-2}.
\begin{figure}[]
	\centering
	\begin{subfigure}{\linewidth} 
	\includegraphics[width=\linewidth]{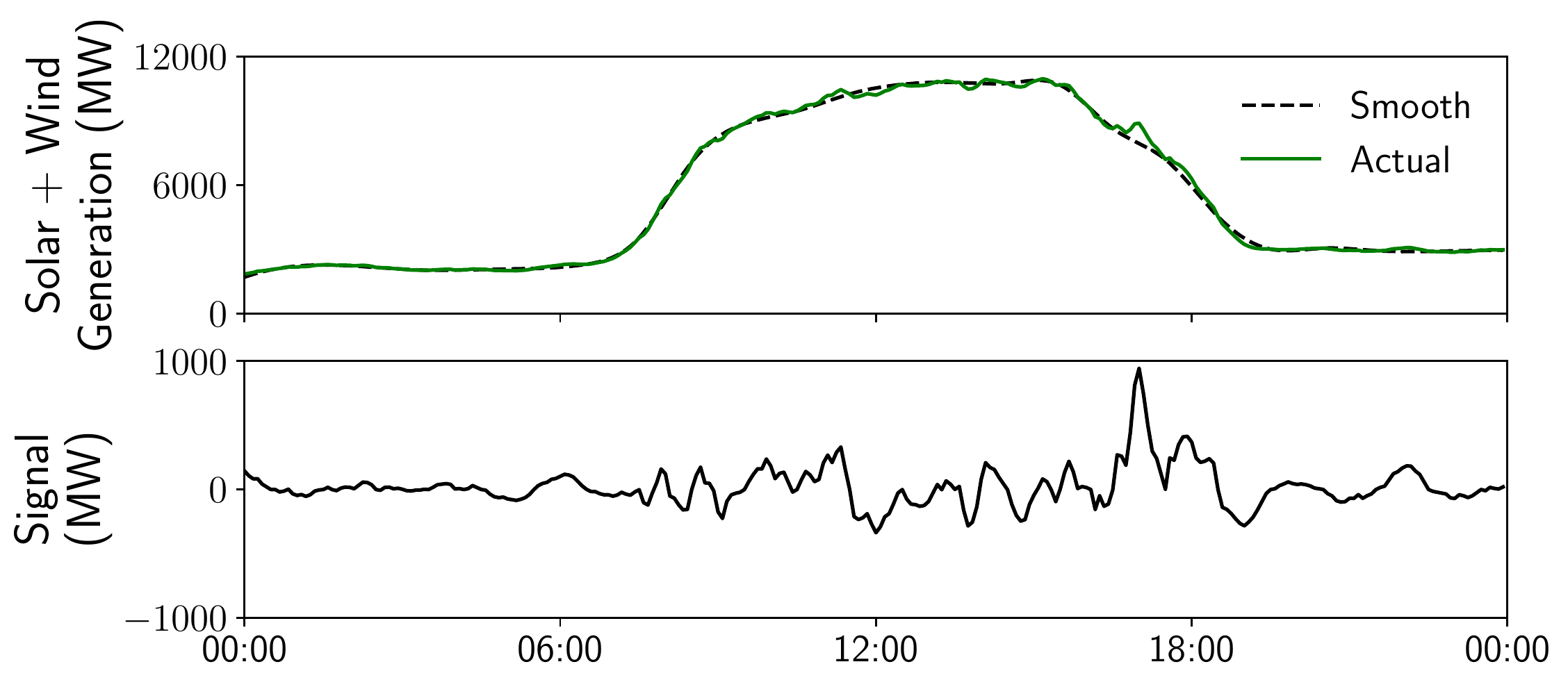}
	\caption{Use Case 1: The 5-min generation following signal is produced by taking the difference between the actual renewable generation and its smooth spline fit, following \cite{burger2017generation}.} \label{fig:ref_signal}
	\end{subfigure}
	\begin{subfigure}{\linewidth} 
	\includegraphics[width=\linewidth]{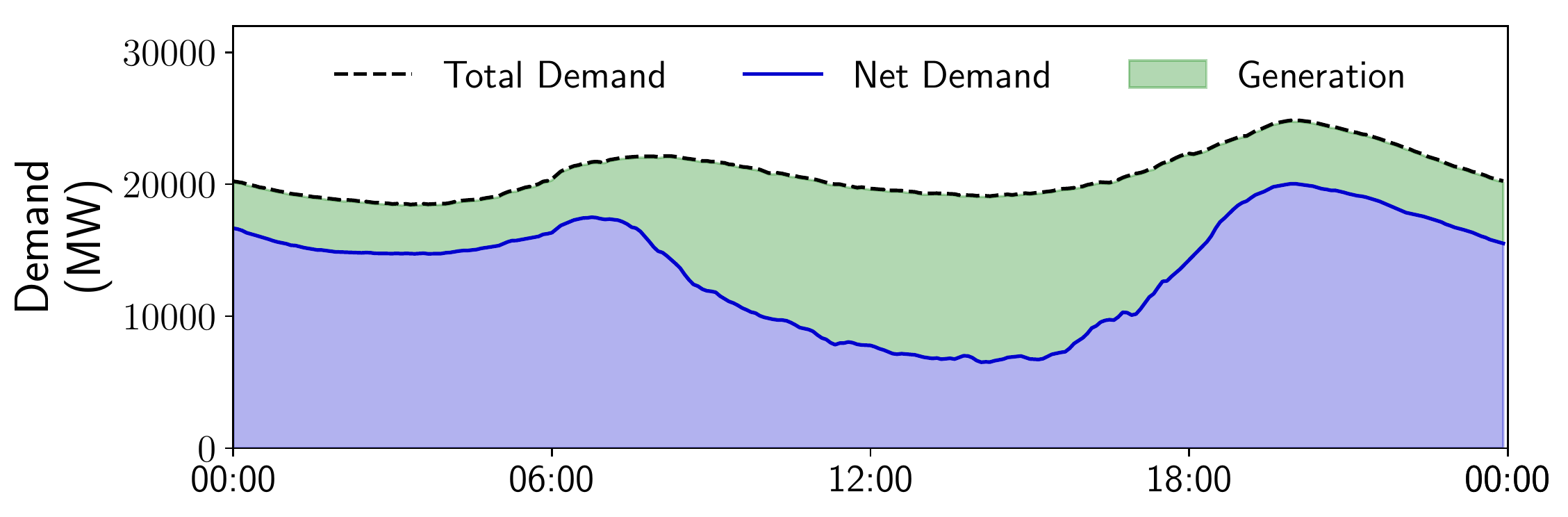}
	\caption{Use Case 2: The duck curve (i.e. the net demand) exemplifies the need for generators to quickly ramp up energy production when the sun sets in areas of high renewable penetration \cite{duckcurve2017}.} 
	\label{fig:duck_curve}
	\end{subfigure}
	\caption{Load Profiles from CAISO, 2020/03/31} 
	\label{fig:load_curves}
\end{figure}

We applied \oursystem~ to address challenges arising from increasing penetration of renewable generation. Firstly, we used the inherent flexibility in the TCL population to absorb the variations in renewable generation in a generation following use case (Section \ref{sec:gen_follow}). Secondly, we shifted the TCL load to alleviate the need to quickly ramp up / down energy generation in areas of high renewable penetration (Section \ref{sec:duck_curve}). The load curves used for both use cases (Figure \ref{fig:load_curves}) are from CAISO \cite{CAISO2020} on 31 March 2020.

\begin{figure*}[]
    \centering
    \includegraphics[width = \textwidth]{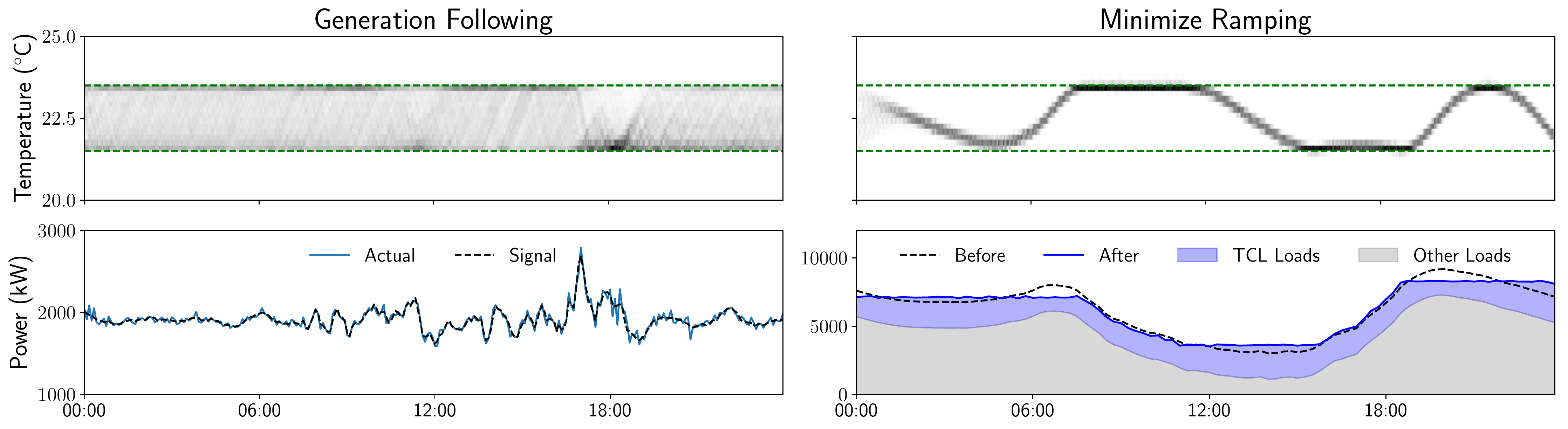}
    \caption{Simulation Study. (Top) the temperature distribution and (Bottom) the aggregate power of the population}
    \label{fig:sim_result}
\end{figure*}

\subsection{Use Case 1: Generation 
Following}\label{sec:gen_follow}
The generation following signal was produced following the same procedure in \cite{burger2017generation}, as shown in Figure \ref{fig:ref_signal}. The TCL population tracked a scaled version of the generation following signal around its baseline power consumption. The objective function is the mean squared error (MSE) between the reference signal, denoted by $\mathbf{\tilde{u}}$, and actual aggregate energy consumption (Eq. \ref{eq:obj_track}). Other tasks such as 
DR events \cite{zhang2013aggregated, tindemans2015decentralized} and frequency regulation \cite{hao2014aggregate, zhao2017geometric} boil down to tracking a given reference signal, and thus may be addressed with the same objective function.

All the optimization problems in this work\footnote{The code is available at \url{https://github.com/INFERLab/COHORT}.} were solved using \texttt{CVXPY} \cite{diamond2016cvxpy} with hyperparameter $\rho=10$. In this use case, we used a 5-min control time-step, and planned for the next time-step, i.e., T = 1. For tracking with PWM, it is necessary to use a smaller time-step. Throughout this work, we used a tracking time-step that is 1/15 of the control time-step. The error limit used for $\Sigma$-$\Delta$ modulation is 0.1kWh in the simulation study.

\begin{equation}\label{eq:obj_track}
g_{\text{track}} \big(\sum_i \mathbf{u_i}\big) = \frac{1}{T}|| \mathbf{\tilde{u}}- \sum_i \mathbf{u_i} ||^2_2
\end{equation}

The behavior of the TCL population is shown in Figure \ref{fig:sim_result}. \oursystem~tracked the reference signal with a small error, while maintaining the temperature of the population within the deadband (dashed green line). Note that the discrepancy between the reference signal and actual power consumption came solely from discretization error. The performance of our approach with comparison to seminal works on TCL control is summarized in Table \ref{tab:track_results}. Our tracking performance is comparable to that in \cite{koch2011modeling}, reported in normalized\footnote{by average aggregate power} root MSE (RMSE), 
and is not as good as \cite{hao2014aggregate}, reported in mean absolute percentage error (MAPE). 
While \cite{hao2014aggregate} may be more suited for reference tracking tasks, it is not applicable to any planning-based task, e.g. load shifting.
Compared to the baseline scenario where the TCL population only maintains temperature, our approach increased the switching frequency by 158.7\%, which is similar to \cite{hao2014aggregate, koch2011modeling}. By adjusting the error limits in $\Sigma$-$\Delta$ modulation, one can trade-off tracking performance and switching frequency. 

\begin{figure}
    \centering
    \includegraphics[width = \linewidth]{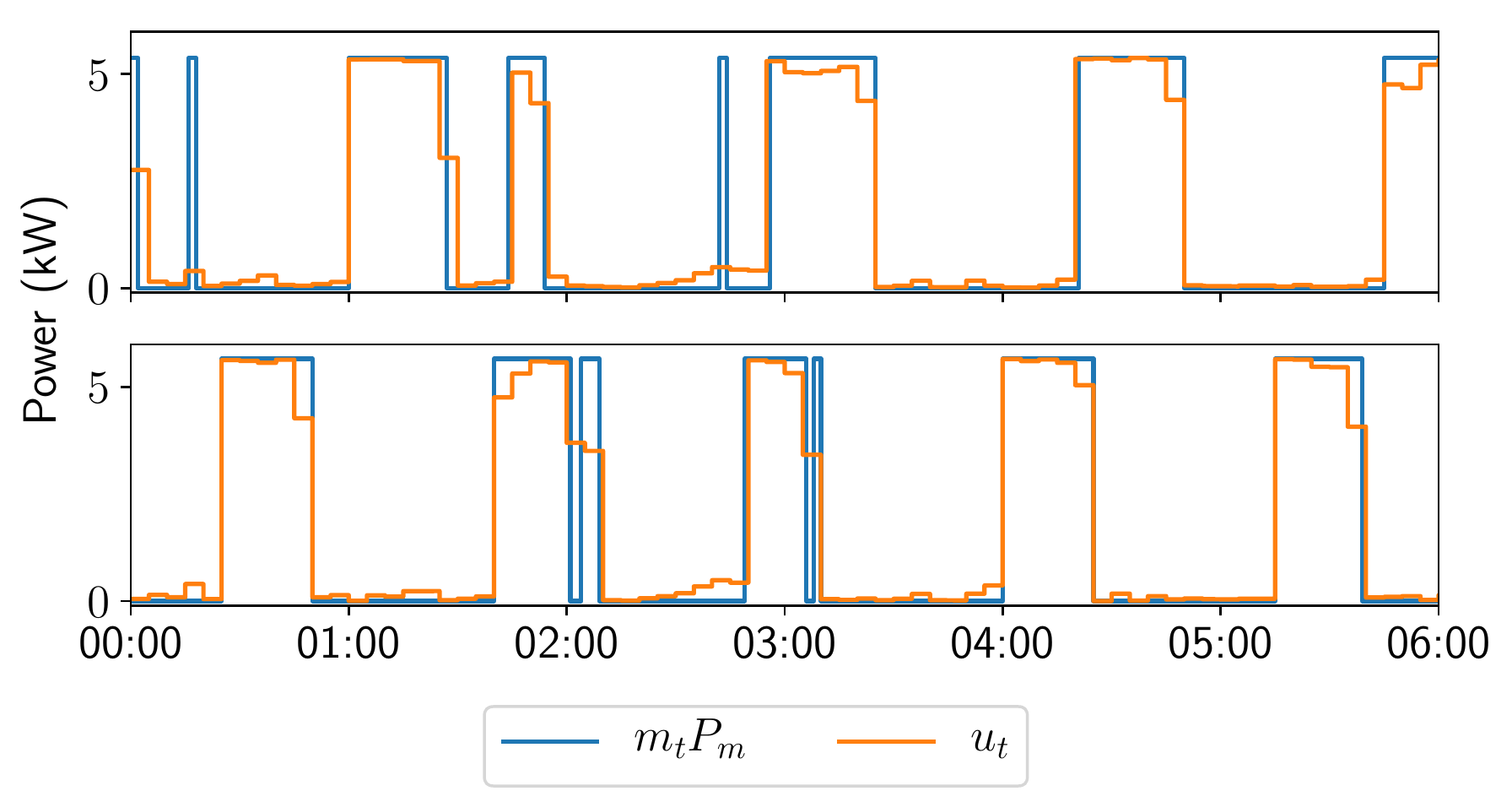}
    \caption{Actions of Individual TCLs}
    \label{fig:pwm_example}
\end{figure}

Figure \ref{fig:pwm_example} show actions of individual TCLs. We initialized each action with either 0 or $P_m$ based on its previous action, and switched if the temperature was close to the edge of the deadband. Given this initialization scheme, the majority of continuous actions $u_t$ are close to either 0 or $P_m$, making conversion to on/off actuation possible with a reasonable number of switchings. 

In this use case, our approach took an average of 5.4 iterations to reach consensus. Interestingly, the number of iterations till convergence is almost independent of the population size, which is also observed in \cite{gebbranpractical}. Since the \textit{u-update} step is distributed to and computed in parallel at each TCL, the overall computation time and the computation cost at the load aggregator scale very well with the population size.

\begin{table}[t]
\caption{Performance Comparison for Reference Tracking}
\begin{tabular}{c c c  c}
\hline  
& \multicolumn{2}{c}{\textbf{Tracking}}&\multirow{2}{1.4cm}{\centering \textbf{Switching Increase}}\\
& \textbf{Norm. RMSE} & \textbf{MAPE} & \\\cline{2-4}
&\textbf{(\%)}&\textbf{(\%)}&\textbf{(\%)}\\
\hline
\citep{koch2011modeling}& 0.8-2.27& - &170-300\\
\cite{hao2014aggregate}& - &<1&116.7\\\hline
\textbf{\oursystem}&2.04&1.53& 158.7\\
\hline
\end{tabular}
\label{tab:track_results}
\end{table}

\subsection{Use Case 2: Minimize Ramping} \label{sec:duck_curve}

While we first evaluated our approach in a reference tracking use case, a major advantage of \oursystem~compared with existing methods is its ability to coordinate TCLs over long time horizons. Thus, we applied our approach to flatten the \textit{duck curve} by shifting load over a day. Specifically, we used a 15-min control time-step, and plan for an entire day ahead, i.e., T= 96. An example of the \textit{duck curve}, named after its resemblance to a duck \cite{duckcurve2017}, is given in Figure \ref{fig:duck_curve}. We scaled down the load curve such that the TCL demand accounts for 20\% of the total demand \cite{hao2014aggregate}. We formulate the objective as minimizing total ramping, i.e., the difference in net demand between consecutive time-steps (Eq. \ref{eq:obj_tv}). $\mathbf{P}_{\text{total}}$, $\mathbf{P}_{\text{net}}$, and $\mathbf{P}_{\text{gen}}$ are the total demand, net demand, and renewable generation, respectively. This problem is also known as total variation minimization \cite{condat2013direct}. While this objective is trickier to optimize, convergence to global optimum is still guaranteed, because $g_{\text{tv}}$ is a convex function. 
\begin{equation}\label{eq:obj_tv}
\begin{aligned}
g_{\text{tv}}\big(\sum_i \mathbf{u_i}\big) &= \sum_{k=t+1}^{t+T}|P_{\text{net,k}} - P_{\text{net,k-1}} |\\
\text{where,}\quad \quad \mathbf{P}_{\text{net}} &= \mathbf{P}_\text{{total}}-\mathbf{P}_\text{{gen}}\\
\mathbf{P}_\text{{total}} &=  \mathbf{P}_{\text{non-shiftable}} + \sum_i \mathbf{u_i}
\end{aligned}
\end{equation}

The behavior of the TCL population for this use case is also shown in Figure 3. The TCLs systematically shifted their load and reduced ramping by 23.1\% compared to the baseline scenario, where the TCL population was operated by on-off control. The temperature of the TCLs shifted accordingly within the deadband. Since the TCLs are operating in cooling mode, reduced energy consumption results in higher temperature, and vice versa. Note that the 23.1\% reduction in ramping is based on the assumption that 20\% of total demand is flexible. Further reduction is possible by extending our approach to other flexible loads.

\section{Experiment 2: Hardware-in-the-loop Simulation}\label{sec:exp-2}

 In this section, we validated \oursystem~ in a HIL simulation, with primary focus on its performance on a real-world testbed. Specifically, we controlled a residential AC via a smart thermostat. More details on the real-world testbed is presented in Section \ref{sec:myApartment}. We augmented the testbed with simulated instances of residential ACs, modeled after real-world data traces (Section \ref{sec:ecobee_model}). We integrated the real-world system and the simulated instances as a HIL simulation, and showcased it in a peak load curtailment use case, based on load profiles from PJM \cite{PJM2018}, as elaborated in Section \ref{sec:hil_setup}. The HIL simulation was executed from 11-25 July 2020, a 15-day period, and the results are summarized in Section \ref{sec:hil_results}.


\subsection{Real-World Testbed}\label{sec:myApartment}
\begin{figure}[]
    \centering
    \includegraphics[width = \linewidth]{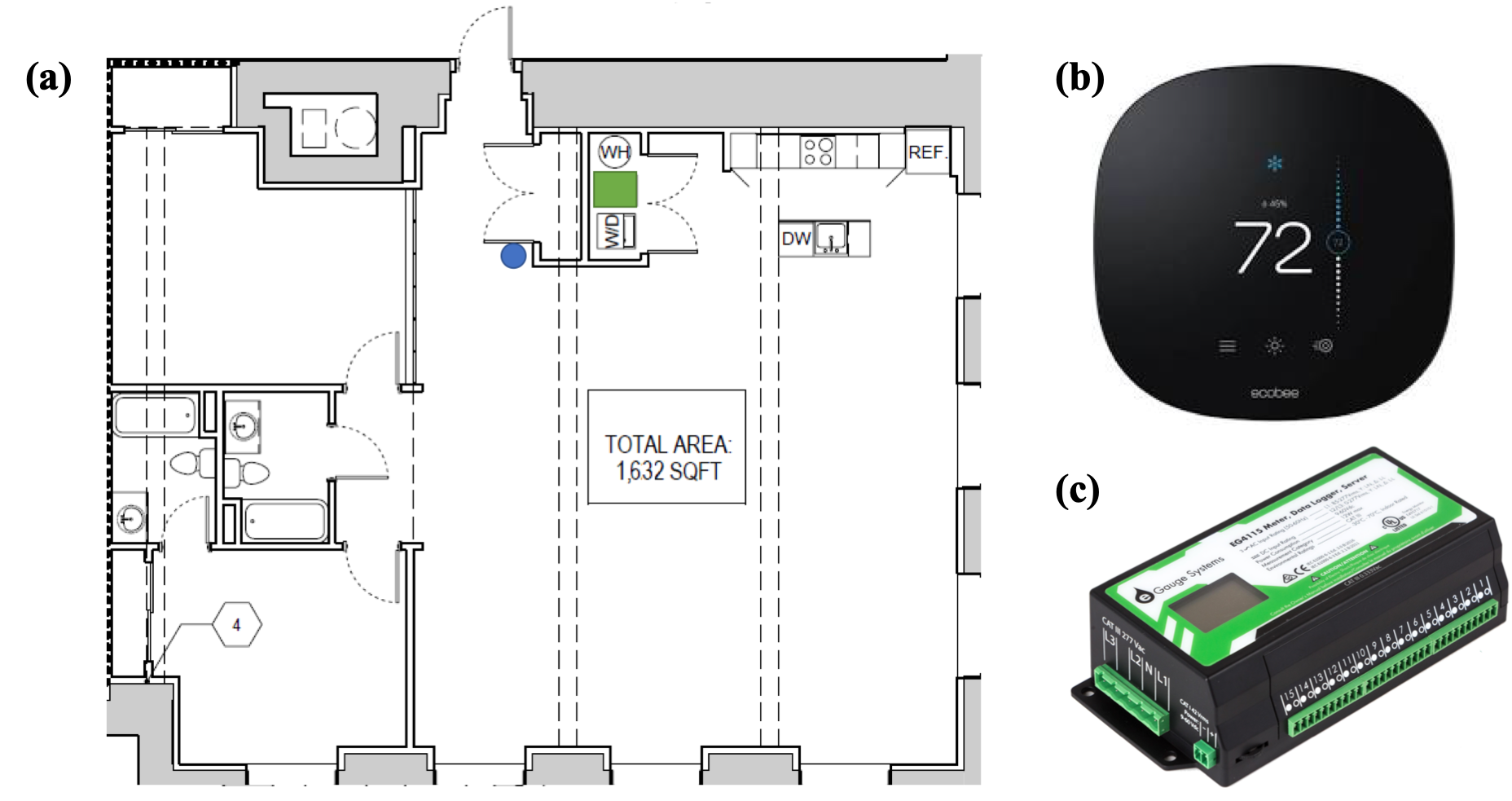}
    \caption{The floor plan of the apartment is shown in (a), with the circle and the rectangle marking the location of the smart thermostat and the indoor AC unit. The apartment is instrumented with (b) an ecobee smart thermostat and (c) an eGauge energy metering unit.}
    \label{fig:myApartment}
\end{figure}

The testbed is first author's apartment located in Pittsburgh, PA, USA (Figure \ref{fig:myApartment}). The 1632 ft\textsuperscript{2} apartment has three regular occupants and was occupied most of the time during the experimental period. The AC unit is controlled via an \texttt{ecobee} smart thermostat, installed at a location shaded from direct solar radiation, and energy consumption of the AC is monitored for verification only via \texttt{eGauge} energy metering \cite{egauge}, with sampling rate up to 1Hz. Power measurements are not needed for control. 

\begin{figure}
    \centering
    \includegraphics[width = \linewidth]{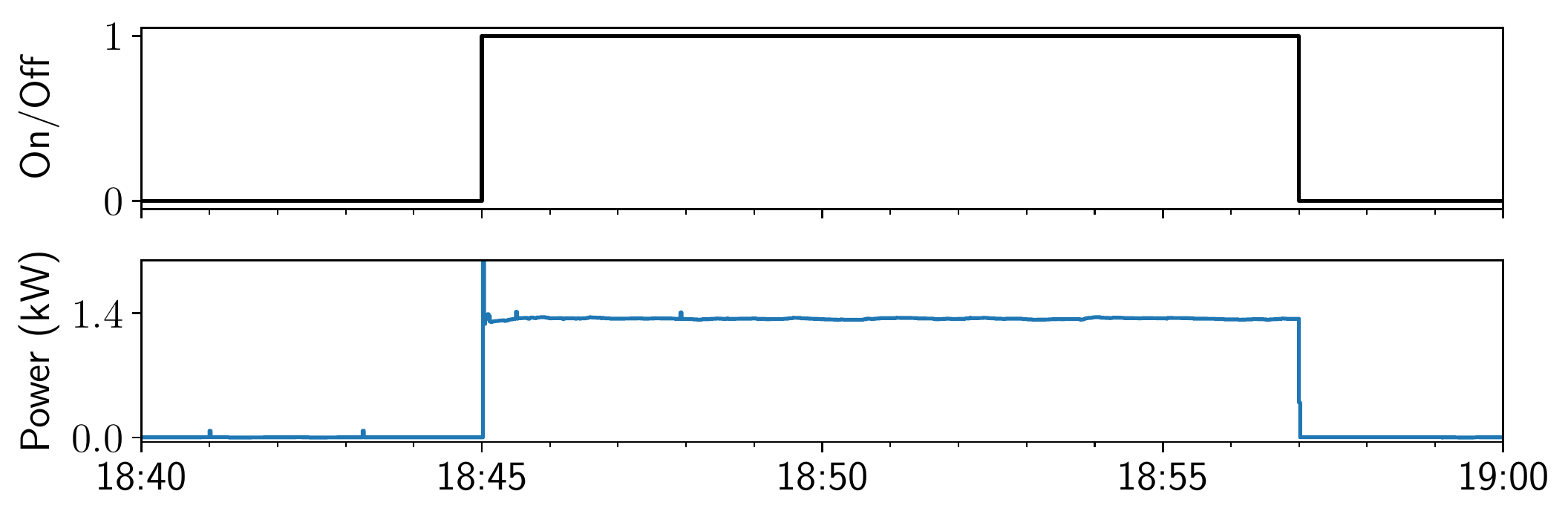}
    \caption{Comparison of the on/off commands vs. the actual power draw by the AC measured at 1Hz}
    \label{fig:nice_match}
\end{figure}

We monitored zone temperate and sent commands to the smart thermostat via ecobee API \cite{ecobeeAPI}. To get the effect of on/off commands, we sent a low temperature setpoint (70 \textsuperscript{o}F) when we want the AC to be \textit{on} and a high temperature setpoint (80 \textsuperscript{o}F) when we want the AC to be \textit{off}. This simple strategy worked surprisingly well. Figure \ref{fig:nice_match} shows a comparison of the commands vs. the power draw on the AC circuit (at 1Hz) for an on/off event. The response of the AC to \textit{on} is almost instantaneous, and the response to \textit{off} is delayed by a few seconds, but negligible compared to the control time-step. Such response was observed throughout the experiment. The ease of integration with a smart thermostat implies \oursystem~could be scaled to 11\% of households in the US already equipped with smart thermostats \cite{king2018energy}, with minimal effort and no retrofit. 

\subsection{Modeling of Residential ACs}\label{sec:ecobee_model}

To have a population of simulated ACs with realistic thermodynamics and population heterogeneity, we took advantage of ecobee's \textit{Donate Your Data} dataset \cite{DYD}. For a comprehensive description of the dataset, we refer interested readers to \cite{huchuk2018longitudinal}. We selected households in the same municipal area as the real-world testbed using data from 2019. We only used households with single-stage cooling, no less than 60 cooling days\footnote{Cooling days are defined as days the AC is operating exclusively in cooling mode, following \cite{huchuk2018longitudinal}.}, and less than 10\% missing data. These criteria left us with 106 households. While the simulated population size is relatively small, the primary focus of the HIL simulation is on the real-world testbed.

The raw data came in 5-min intervals and we down-sampled it to 15-min based on the control time-step. We used the same model form as \cite{kara2014quantifying} (Eq. \ref{eq:arx}), where $T_t$ is the control temperature, $u_t$ is the duty cycle ratio, $T_{a,t}$ is the outdoor ambient temperature, and $d_{o,t}$ and $d_{s,t}$ are binary flags for occupancy sensor activation and scheduled sleep time. These variables would be explained shortly. Both the model orders, i.e., $p$ and $q$, and the disturbance terms were selected based on the Akaike Information Criterion (AIC). The median of selected model orders are $p=5$ and $q=2$. 
\begin{equation}\label{eq:arx}
    T_{t+1}= \sum_{i=0}^{p-1} a_i T_{t-i} + \sum_{i=0}^{q-1} b_{u,i} u_{t-i} + b_{a}T_{a, t} + b_{o} d_{o, t}+b_{s} d_{s,t}
\end{equation}

 The state variable, the \textit{control temperature}, is what an ecobee uses for operating the AC with respect to the setpoint. It is a weighted average of the temperature measurements at the main thermostat and remote sensors \cite{huchuk2018longitudinal}. Note that the control temperature in the dataset came in 1\degree F resolution. The control action, \textit{duty cycle ratio}, is the equipment run-time normalized between 0 and 1. Similar to \cite{huchuk2018longitudinal}, we assumed that the house was occupied if any of the motion sensors were triggered or during scheduled sleep time. Since the raw data from occupancy sensors were sharp spikes, we passed the data through a low-pass filter. We used a separate variable for scheduled sleep time. Despite having different form from Eq. \ref{eq:sys_relaxed}, Eq. \ref{eq:arx} can also be written in the form of Eq. \ref{eq:vb_matrix} as a linear system,
 and thus the same formulation of flexibility (Eq. \ref{eq:flex}) still applies.

\begin{figure}
    \centering
    \includegraphics[width = \linewidth]{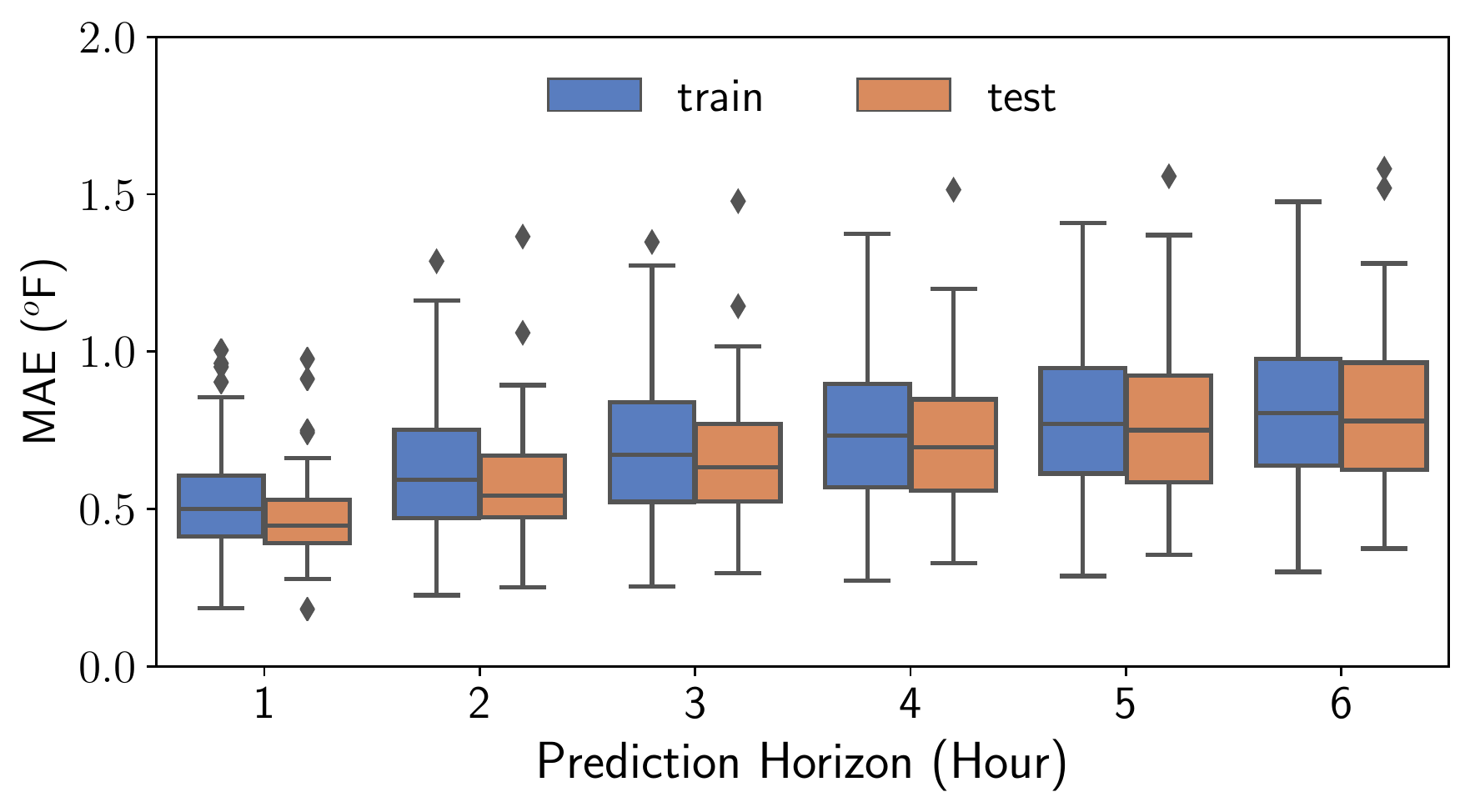}
    \caption{Distribution of model prediction error for 1 to 6 hour prediction horizons over 106 households}
    \label{fig:mae}
\end{figure}

We evaluated the modeling performance with mean absolute error (MAE) over 1 to 6 hour prediction horizons. Figure \ref{fig:mae} shows the distribution of MAE over 106 households. The train set and the test set were the first 2/3 and last 1/3 of cooling days, respectively. The weather was checked to have a similar distribution over the train-test split. The majority of prediction error is less than 1\degree F even at a 6-hour prediction horizon. This result is comparable to that of \cite{kara2014quantifying}. Bear in mind, in interpreting the results, that the control temperature came in a low resolution of 1\degree F. 

\subsection{Experimental Setup}\label{sec:hil_setup}
Growing peak demand decreases the average utilization of generators  \cite{eiapeak} and increases the need to build and operate high marginal cost peaking generation \cite{callaway2010achieving}. Thus, we showcased our approach on a peak load curtailment use case, the objective of which is given in Eq. \ref{eq:obj_peak}. As an infinity norm\footnote{Infinity norm is defined as $||\mathbf{x}||_\infty := \max(|x_i|)$.} minimization problem, the solver minimizes the maximum total load within the planning horizon. We used a 15-min control time-step and found a 16-hour planning horizon to be sufficient. We re-planned at each hour to avoid compounding modeling error. Similar to the experiment in Section \ref{sec:duck_curve}, we scaled down the PJM load profile with the assumption that TCL loads account for about 20\% of the total load. 

\begin{equation}\label{eq:obj_peak}
\begin{aligned}
g_{\text{peak}}\big(\sum_i \mathbf{u_i}\big) &=||\mathbf{P}_{\text{total}}||_\infty\\
\text{where,}\quad \quad \mathbf{P}_\text{{total}} &=  \mathbf{P}_{\text{non-shiftable}} + \sum_i \mathbf{u_i} 
\end{aligned}
\end{equation}

We integrated the real-world testbed with simulated instances of residential ACs to form an HIL simulation. The HIL simulation is also integrated with day-ahead weather forecast via Dark Sky API \cite{darksky} and total load profile from PJM Data Miner 2 \cite{PJM2018}. 

We modeled the real-world testbed using the same procedures as described in Section \ref{sec:ecobee_model}. Note that the temperature time series pulled from ecobee API comes in 0.1\degree F resolution, and thus allows for more accurate modeling. To identify the system accurately, the testbed was excited by square wave signals, one-hour on followed by one-hour off, on 10 July 2020. During the experiment, the model was updated regularly based on new data. We used $P_m$=1.4kW for the real-world system based on actual power measurements. As discussed in Section \ref{sec:pwm}, we initialize $\mathbf{u_i}$ with sequences of interlaced 0s and $P_m$s to encourage sparsity in the solution. For the AC unit, we initialized the solution for each hour with [$P_m$, 0, 0, 0] and used an error limit of 0.075kWh. 

For the simulated instances, we selected models that performed no worse than 75\textsuperscript{th} percentile on any of the prediction horizon, which left us with 72 households. The dataset does not contain information on the rated power of the AC units, and thus we sized the AC based on the floor area, following Energy Star recommendations \cite{sizeAC}. We used the same occupancy data from the same time last year. The temperature setpoint for the entire population, including the real-world testbed, was assigned to be 75\degree F throughout the experiment.

\vspace{-.5em}
\subsection{Results}\label{sec:hil_results}
\begin{figure}[]
	\centering
	\begin{subfigure}{\linewidth} 
	\includegraphics[width=\linewidth]{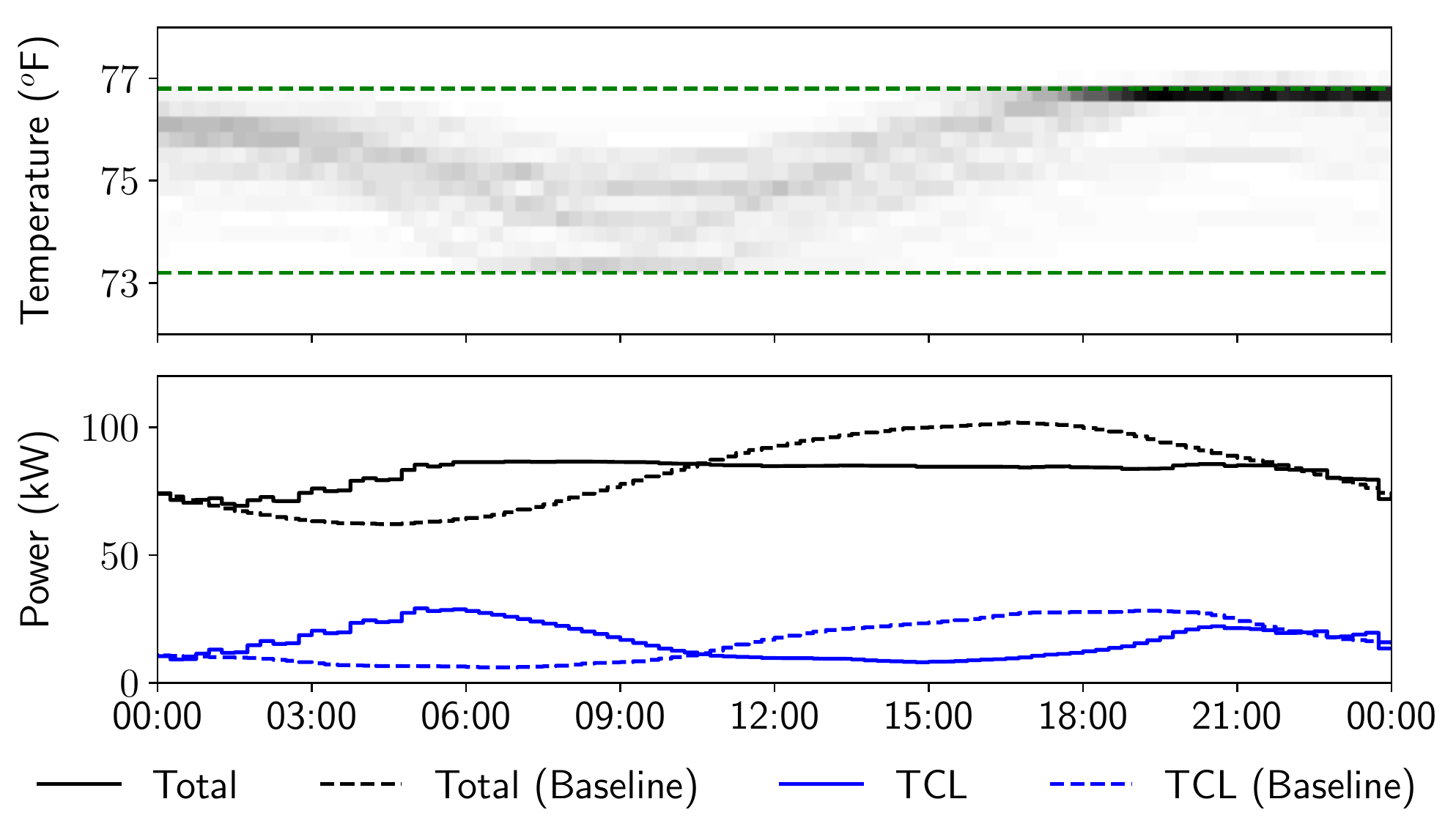}
	\caption{(Top) temperature distribution and (Bottom) aggregate power} \label{fig:agg_ts}
	\end{subfigure}
	\begin{subfigure}{\linewidth} 
	\includegraphics[width=\linewidth]{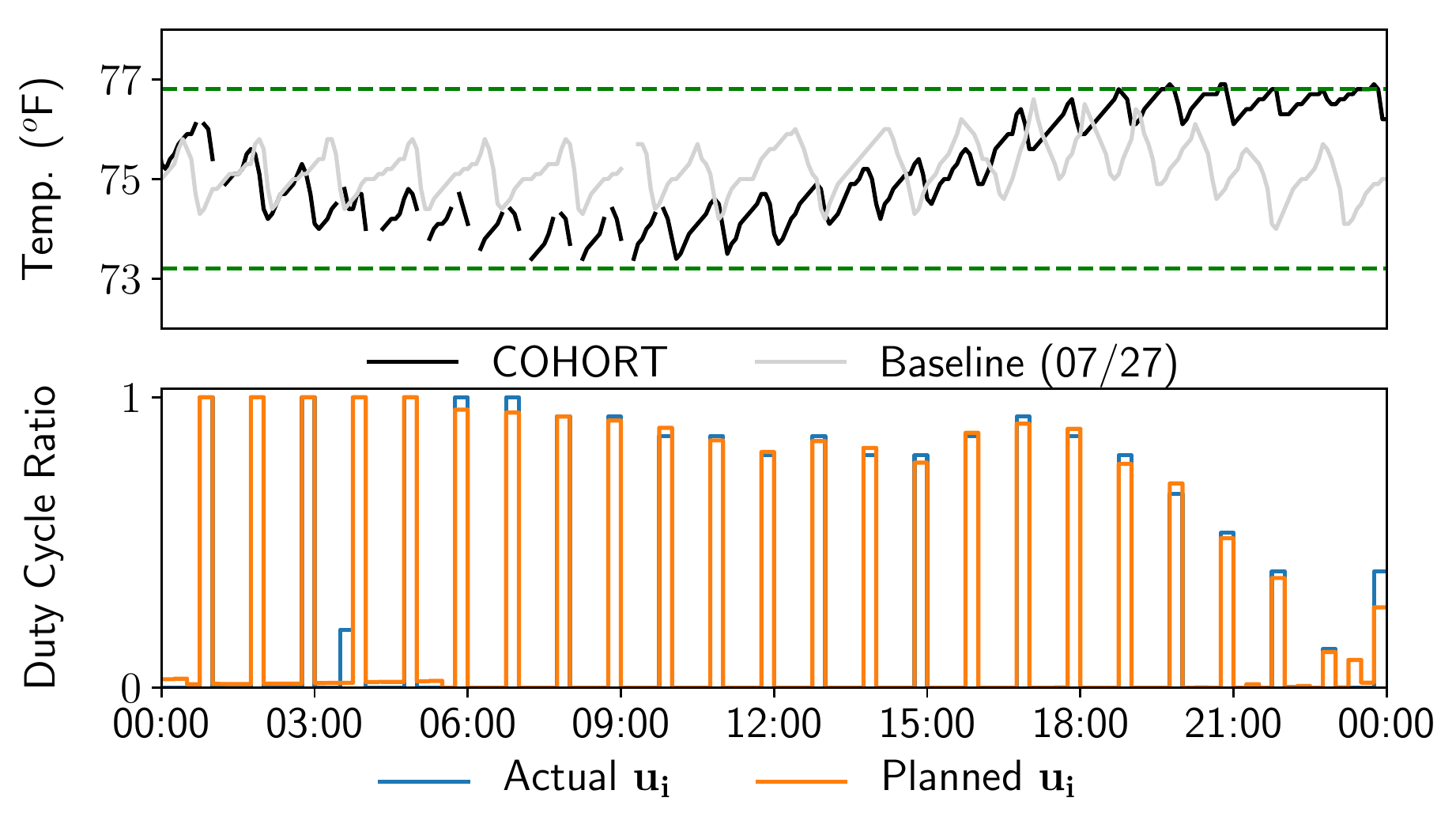}
	\caption{(Top) temperature and (Bottom) duty cycle ratio} 
	\label{fig:individual}
	\end{subfigure}
	\caption{HIL Simulation. Behaviour of (a) the population and (b) the real-world testbed on 2020/07/24 }
	\label{fig:ts}
\end{figure}

The performance of \oursystem  ~compared with the baseline scenario, where the TCL population was operated by on-off control, 
is exemplified in Figure \ref{fig:ts},
using time series from 24 July 2020. The baseline TCL load was simulated using a model of the real-world testbed, along with the rest of the population, based on the same weather data. The baseline total load was a scaled down version of the actual PJM load profile, under the assumption that TCLs account for about 20\% of the total load. At the aggregate level (Figure \ref{fig:agg_ts}),  the TCL peak approximately coincides with the utility peak in the baseline scenario (dashed lines). \oursystem~systematically shifted the TCL load away from peak hours to early morning, thereby reducing the utility peak by 15\%. The temperature of the population shifted accordingly. By pre-cooling the households when the demand was low, the TCLs could reduce energy consumption during the peak hours and let temperature slowly float up, without violating comfort constraints. The same behavior is observed in the real-world testbed (Figure \ref{fig:individual}). 

Also in the real-world testbed, the planned vs. actual duty cycle ratio at each 15-min control time-step shows good overall agreement. By initializing the actions at each hour with [$P_m$, 0, 0, 0], the \textit{on} events were spaced apart, thereby reducing the risk of short cycling. The temperature, as logged by ecobee at 5-min intervals, was maintained within the deadband. To compare with the baseline scenario, we also include in the same plot the temperature time-series in the real-world testbed on 27 July 2020, when the ecobee maintained temperature at 75\degree F setpoint using its default strategy. The slightly wider temperature range and the gradual temperature shift were barely perceptible by the occupants. All occupants were happy with their thermal comfort during the experiment. In single blind tests, occupants not involved in the experiment were not able to tell if the AC was operated by \oursystem~or on-off control.

Over the 15-day experimental period, COHORT reduced daily peak loads by an average of 12.5\% (with a 2.9\% standard deviation), when the TCL accounts for about 20\% of the total load. In comparison, the peak load was reduced by 8.8\% in the best-performing case in \cite{cole2014community}, when the TCL peak account for 74\% of the utility peak in the baseline scenario. While the experimental setups are different, the difference in performance likely comes from the fact that a mere 1-hour planning horizon was used in \cite{cole2014community}. Instead, we planned ahead for 16-hour and managed to level the total load from about 6:00-21:00 in the case shown in Figure \ref{fig:agg_ts}. This affirms the advantage of being able to plan over long time horizons.

\section{Conclusions}\label{sec:discussion}

We proposed \oursystem, a novel distributed control solution for coordinating TCLs to jointly optimize a grid-level objective, while satisfying each TCL's end-use requirements and operational constraints. Our approach decomposes the grid-level problem into subproblems and coordinates their solutions to find the grid-level optimum. To be computationally viable over long planning horizons, we apply convex relaxation to the discrete action space and characterize each TCL's flexibility as a convex set. After coordination, each TCL tracks the agreed-upon power trajectory locally with a PWM-based strategy, which translates the continuous power back to on/off actuation.  
Since each TCL is responsible for its own control, it can incorporate detailed and system-specific dynamics and constraints, which is difficult to accomplish in a centralized architecture. Furthermore, the coordination process is independent of each TCL's dynamics and control scheme, making \oursystem~extensible to other flexible loads. 

We validated \oursystem~in simulation studies and a HIL simulation to address challenges arising from growing peak demand and increasing penetration of renewable generation. In the generation following use case, our approach showed comparable performance to prior work. A major advantage of our approach compared with existing work is its ability to coordinate TCLs over long planning horizons. Thus, we applied it to load shifting use cases, assuming TCLs account for 20\% of the total load \cite{hao2014aggregate}. Firstly, we used it to smooth out the duck curve, based on actual load profile from CAISO, and reduced ramping by 23.1\% compared to the baseline scenario. Secondly, we applied it to curtail peak load based on load profile from PJM. The experiment was conducted on a HIL simulation, including a real-world testbed. Over the 15-day experiment period, \oursystem~ reduced daily peak loads by an average of 12.5\%. Furthermore, the occupants living in the real-world testbed, i.e., the first author's apartment, reported no discomfort and could not distinguish whether the AC was operated by on-off control or \oursystem. Finally, \oursystem~is easily integrated with a commercial smart thermostat, which provides readily-available control and communication infrastructure. Thus, our approach is scalable to households already equipped with smart thermostats with minimal effort and no retrofit.

In summary, \oursystem~ is shown to be a practical, scalable, and versatile solution for coordinating TCLs to provide grid services. Currently, quite a few iterations are required to reach consensus for problems with long planning horizons, which may raise concern about the communication cost. This was discussed in \cite{gebbranpractical}, with the conclusion that the network latency and message size between the aggregator and the end users during coordination are not bottlenecks for modern networks\footnote{More concretely, the message size is less than $1\text{KB}$ per agent per iteration in our problem setup and the total message size grows linearly with the number of agents. Overall, the bandwidth requirement is low. As a reference, the network latency for 4G network ranges from 30 to 160ms. Even for 100 iterations, the latency is small compared to a 1-hour re-planning time-step sufficient for load shifting use cases.}. Regardless, the number of iterations could be reduced by initializing \oursystem~with an approximate solution from imitation learning, to be incorporated as future work. Another research direction would be to extend the current work to other flexible loads, with tracking strategies tailored to those loads. 

\begin{acks}
This material is partly based upon work supported by the U.S. Department of Energy's Office of Energy Efficiency and Renewable Energy (EERE) under the Building Technologies Office Award Number DE-EE0007682. We would like to thank Eric Burger and Scott Moura, whose prior work was the primary inspiration for us. We also thank \texttt{ecobee} and its customers for their data. \textit{Disclaimer: The views expressed in the article do not necessarily represent the views of the U.S. Department of Energy or the United States Government.}
\end{acks}

\bibliographystyle{ACM-Reference-Format}
\bibliography{sample-base}

\end{document}